\documentclass{llncs}
\usepackage{epsfig}
\begin{document}
\title{BREW: A Breakable Web Application for IT-Security Classroom Use}
\author{Christoph Pohl \and Kathrin Schlierkamp \and Hans-Joachim Hof
\\ \email{christoph.pohl0@hm.edu}, \email{kathrin.schlierkamp@hm.edu}, \email{hof@hm.edu}}
\institute{MuSe - Munich IT Security Research Group \\ Department of Computer Sciences and Mathematics\\ Munich University of Applied Sciences\\ Munich, Germany}
\maketitle

\begin{abstract}
This paper presents BREW (Breakable Web Application), a tool for teaching IT Security. BREW's main teaching targets are identification and exploitation of vulnerabilities, using technologies and methodologies for software auditing and testing, and bug detection, fixation, and writing of secure code. Main advantages of BREW include that it is easy to apply in practice, it is a perfect tool to create and retain motivation, it corresponds to the demands of the psychology of learning, and it can be used for a heterogeneous group of students. BREW has been successfully used for teaching IT Security in Germany as well as on an Erasmus Project with international student groups.
\end{abstract}

\section{Introduction}
As a reaction on Snowden's disclosure of the gigantic scope of spying on IT infrastructure by the NSA press coverage on the topic IT security has increased. It is very likely that the number of students interested in attending IT security lectures or even the number of IT security study programs will increase.
An efficient and motivating way to teach IT security is needed, and the lessons taught should be easy to apply on future software engineering projects of the students.
Teaching as many computer science students as possible how to write secure code, is a good way to raise the security level of systems in the future. Nowadays, many security vulnerabilities exist because programmers are not aware of IT security when writing code.
For example, the   OWASP Top Ten project \cite{owasptopten2013} lists the most common security vulnerabilities of web applications.
The most common security vulnerability of this list, called "Injection", can easily be avoided if the programmer is aware of this problem.
Avoiding SQL Injection, the most common injection, requires only using a special function call to the database.
Injection being the commonest vulnerability of web application shows, that there is a great demand for security education with web programmers.

When teaching secure code writing, different approaches could be chosen:
\begin{itemize}
\item The Hacking Approach: students search for vulnerabilities in web applications and exploit these vulnerabilities
\item The Developer Approach: students are tought coding guidelines to avoid vulnerabilities
\end{itemize}
The first approach teaches students to think like a hacker.
This state of mind includes doing things in an unusual way to find vulnerabilities. It is the aim of the hacker to find ways to use the system that the programmer did not consider. Blind spots of developers may lead to vulnerabilities of software.
However, the hacking approach falls short because it is a destructive approach: it tells how to break software, not how to protect software from an attack.

In contrast, the second approach is a constructive approach: rules are presented, that help to write secure code, hence reduce the number of vulnerabilities that will exist in software.
However, a rules based approach does not support a deeper understanding of IT security and it does not help students to understand how attacks work.
Especially, students do not learn to think "outside the box" like an attacker would do.
Both approaches can be applied to lab sessions: hacking labs or secure software development labs.
However, most teaching tools available today do not offer the combination of both approaches. 

BREW (Breakable Web Application) implements both approaches and hence is an ideal tool to teach IT security.

This paper is structured as follows: The next section gives an overview of BREW.
Section \ref{sec:rel} compares BREW to other tools for teaching security of web applications.
Section \ref{sec:eddesign} presents the educational design of BREW.
The section \ref{sec:tecdesign} shows the technical design as well as some usage scenarios, followed by the section \ref{sec:deploy} with different deployment targets.
Section \ref{sec:exp} summarises some experiences with BREW from classroom use at Munich University of Applied Sciences and the Erasmus Intensive Learning Program "Secure Web Applications".
Section \ref{sec:conclusion} concludes the paper and gives an outlook on work in the future.
 
\section{Overview}
\begin{figure}[h!]
\centering
\includegraphics[width=1\textwidth]{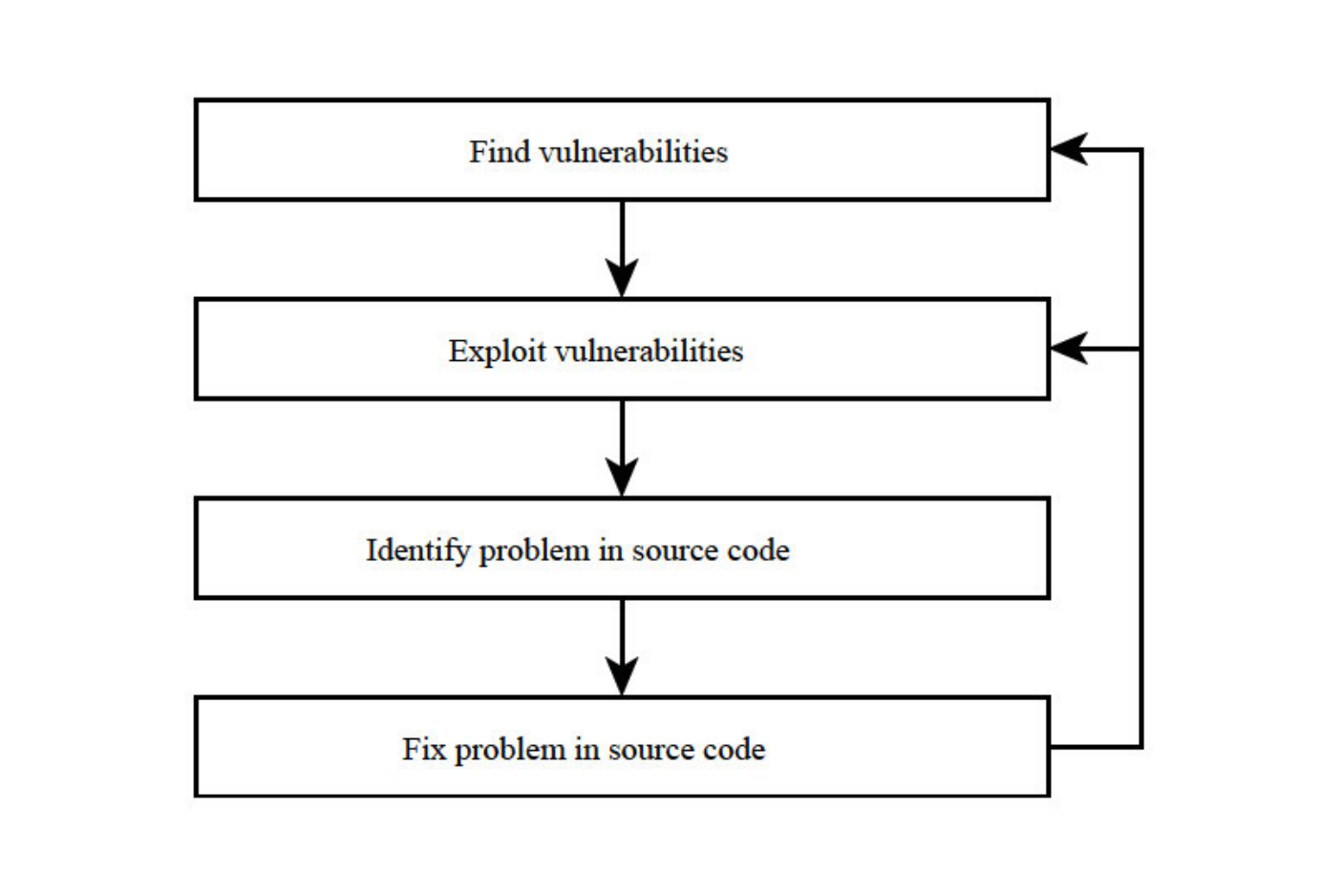}
\caption{Student workflow} 
\label{fig:overview}
\end{figure}

BREW is a platform for hands-on teaching of writing secure web applications.
The main idea of BREW is to have a web application with many security vulnerabilities.
Following the topics of the associated lecture, students first try to find one or more vulnerabilities of the web application.
Each identified vulnerability should then be exploited by the students.
Following this hacking part of a lab, the students change to the source code of the application.
They are asked to identify the lines of code, that introduce the vulnerabilities they exploited before.
Then the students have to fix the vulnerability.
At the end of a web session, the students try to hack the fixed application again and hopefully they are not successful.
Figure \ref{fig:overview} gives a schematic overview for this process.

To summarise, BREW is designed with the three educational targets given in table \ref{tab:edtarget}.

\begin{table}
\centering
\caption{Educational targets}
\label{tab:edtarget}
\begin{tabular}{llllll}
\hline\noalign{\smallskip}
Identification & Target\\
\noalign{\smallskip}
\hline
\noalign{\smallskip}
T1 & Identification and exploitation of vulnerabilities\\
T2 & Technologies and methodologies for software auditing and testing\\
T3 & Bug detection, fixation and write secure software\\
\hline
\end{tabular}
\end{table} 

To be able to use BREW in labs with heterogeneous hardware, it is realised as a virtual machine image in OVA format. OVA can be used in virtualization environments like VirtualBox \cite{vbox2014} (client) or qemu-KVM \cite{kvm2014}. The virtualization approach makes BREW platform independent as VirtualBox is available on a wide range of operation systems, including Windows, MacOS, and Linux. The technical design of BREW allows it to run as a local virtual machine (e.g. on a lab computer or on the laptop of students) or even in a cloud environment like CloudStack \cite{cloudstack2014}, OpenStack \cite{openstack2014} or Amazon Cloud \cite{amazon2014} (not full featured yet).

Inside the virtual machine, BREW is realised as an Integrated Development Environment (IDE) project. Students start the web server by a "play" button of the IDE projects. They connect to the web application by the installed browser. Hacking is done in the browser or using external tools also installed in the virtual machine. Working with code takes place in the IDE.\\

The most important goals of BREW is to teach useful secure code writing for real world software engineering projects. To do so, it is very important that the selection of vulnerabilities covered by BREW is relevant to real world software projects. We chose to base the selection of vulnerabilities on the OWASP Top Ten project \cite{owasptopten2013}. The OWASP Top Ten project lists the most common attacks on web applications. The current top ten list is from the year 2013, so it is quite new. Table \ref{tab:owasptopten} lists the current top ten list. BREW includes vulnerabilities that enable the attacks {\em A1-A3}, {\em A5 - A8}, and {\em A10} from this list.

\begin{table}
\centering
\caption{Owasp Top Ten Attacks}
\label{tab:owasptopten}
\begin{tabular}{llllll}
\hline\noalign{\smallskip}
Identification & Attack type\\
\noalign{\smallskip}
\hline
\noalign{\smallskip}
A1 & Injection attacks\\
A2 & Broken Authentication and Session Management\\
A3 & Cross Site Scripting\\
A4 & Insecure Direct Object References\\
A5 & Security Misconfiguration\\
A6 & Sensitive Data Exposure\\
A7 & Missing Function Level Access Control\\
A8 & Cross Site Request Forgery\\
A9 & Components with Known Vulnerabilities\\
A10 & Unvalidated Redirects and Forwards\\
\hline
\end{tabular}
\end{table}

To deal with the heterogeneous background of students regarding knowledge of relevant technologies as well as to allow a personal learning process, four different stages of complexity of vulnerabilities exist in BREW. BREW implies, that a student with basic knowledge of Java and knowledge from the associated lecture can deal with the first three stages. The fourth stage is mentioned as special challenges and is used for further self educational of the students. More details on stages is given in section \ref{subsec:stages}.

\section{Related Work}\label{sec:rel}
One of the best known platforms for lectures in web application security is \mbox{WebGoat\cite{webgoat2014}.}
Based on java and J2EE it provides different lessons  and exercises related to it-security.
It consists of different lessons exploitable for students.
It aids students with an inbuilt help system and provides solutions for the exercises.
However WebGoat is suitable for the most lectures, depending on black box testing.
The main target relays on detecting vulnerabilities.
The fixation or closing of flaws is not in scope of WebGoat.
This is one of the main differences to BREW.

Most similar to BREW is the Damn Vulnerable Web Application \cite{dvwa2014}.
It provides different stages in complexity (like BREW) and gives hints for the exploitable parts.
It depends on php and mysql and is also available as virtual machine.
It provides black box and white box testing features.
However, the application has only a small amount of vulnerabilities, and the educational concept seems enhanceable.
There is no explanation for the different vulnerabilities or the solution.
And unlike BREW it does not have the look and feel of a normal web application.

An online solution is presented in Hackthissite \cite{hackthissite2014}.
It consists of different exercises with different stages of complexity.
While the source code is not available it is only suitable for black box testing.
The fixation of the vulnerabilities is out of scope from this approach.

\section{Educational Design}\label{sec:eddesign}
This section describes the educational concept behind BREW in IT-Security lectures. 
The educational concept behind BREW offers four major advantages: it is easy to apply in practice, it is a perfect tool to create and retain motivation, it corresponds to the demands of the psychology of learning and it can be used for a heterogeneous group of students. 

Since BREW is an open source application and there is no need for further installation or configuration, BREW circumvents any obstacles that students and/or lectures could have. BREW can be easily explained to lecturers, tutors and students and its system is immediately applicable.
There is no need for specific hardware or software and it can be used for groups as well as for single students.
In sum: after a short introduction BREW is ready to use for every kind of IT-Security lecture.
From an economic perspective BREW satisfies the most important requirements: time and money.
But what about the advantages of the concept behind BREW when it comes to the motivation of students?
There are many different theories and ideas about the way in which lectures should be designed that students can learn as much as possible.
One of the most important and even more often unconsidered fact is motivation.
The biggest challenge for the lecturer is to motivate the students to an extent that they are willing to learn and willing to throw themselves into complex subjects, even if they are struggling \cite{Kluver:2012fs,Lipowsky:ue}.
But how can you create and retain motivation during a semester?
And moreover: how can you build an educational design that creates motivation and covers the approached content of the lecture?
According to the ARCS model designed by Keller there are four important elements that a lecture should generate in every single student: Attention, relevance, confidence and satisfaction \cite{Hubwieser:2007vs,Keller:1987we,Keller:1987uo}.
Attention means that a lecture must contain interesting examples, mental challenges and information which are in a personal and/or emotional way important for the student.
Relevance can be provided by revealing the intended goal of the lecture.
Moreover the lecture should inspire the student with confidence the student should know that even if the lecture is sometimes hard to understand or to put in practice, at the end he or she will manage all difficulties and rise to the challenge.
And last, there should be satisfaction.
Every learning cycle should offer the possibility to achieve a goal.
Therefore the learning cycle must be adoptable to the different skill levels of every attending student.
 
BREW contains all four elements.
First, it is a perfect and most realistic example for vulnerabilities every major company has to face with.
Students can get a realistic insight what secure software has to supply and which are the most popular problems it has to strive against.
Second, right at the beginning the three major targets are identified: the identification and exploitation of vulnerabilities, the knowledge of technologies and methodologies for software auditing and testing, further the bug detection, fixation and writing of secure software.
Third, BREW offers a manual which makes it possible to offer immediate help if the student is struggling with some problems.
Due to the fact that BREW works with different skill levels, in the fourth place every student has the chance to solve successfully the given tasks.
Moreover, students experience the challenge for further research or knowledge.
As you can see, BREW is the ideal way to provide a lecture that can create and retain the motivation for (almost) every student who is interested in the field of IT security.
But BREW has some more educational advantages to offer.

According to the psychology of learning, a successful learning behaviour depends not only on the individual learning rate and strategy of a student, but also on the way the content of a lecture is presented to the students.
Therefore the lecture must be structured according to the following three principles.
First, a logical design and a logical chain of thought; for example a connection between different kind of problems or dependencies among software elements.
Second, a specific user experiment which appeals to all senses.
Third, a lecture should be earmarked to a specific purpose.
The whole structure of a lecture must therefore reflect the learning target \cite{Brocke:2007eg,Mienert:2011kx,Wild:2009uz}. 
BREW can also cope with those three requirements.
All challenges are built systematically on one another.
By using BREW students have a one on one real-time user experiment: the way in which BREW works is exactly the same way every secure software works and it represents for example the bug fixation realistically.
Thus BREW reflects the learning target on all levels in a perfectly matched way.
Therefore BREW is the ideal instrument to face one of the biggest challenges - the heterogeneous groups of students with different skill levels.
 
Over the recent years lecturers experienced that an increasing number of students show difficulties in the acquiring of software knowledge as well as in coping with the study process itself.
One reason for this negative trend can be found in the significant deficiencies in certain base competencies like self-, practical or cognitive competencies that are relevant for successfully studying software related topics \cite{Bottcher:2014wv}. 
In order to face and overcome those difficulties, lecturers must seek different kinds of Software Engineering education tools.
Fortunately BREW could be one of those required tools.

BREW challenges a wide range of the most important competencies for the acquiring of software knowledge like reading skills, systematic and methodical acting, understanding of written instructions, abstract thinking as well as acting in a goal-oriented way.
Moreover, students who are using BREW exceed the learning of pure knowledge but rather learn the active application.
Therefore BREW gives the opportunity to cover the deficits in those skills in a motivating and challenging way for students with different competency levels.\\ 

\section{Technical Design and Usage Scenarios}\label{sec:tecdesign}
This section describes the technical design in detail. BREW is quite flexible, can be adapted and is useful in many scenarios. This sections shows some common usage scenarios.

\subsection{Stage Design}\label{subsec:stages}
To deal with different technical backgrounds of students as well as with lecture process and to keep hacking interesting for the students, BREW offers stages with different complexity. The higher the complexity, the harder it is either to identify or to exploit a vulnerability. The first three stages are used for common labs. The fourth stage is very difficult to master, hence remains an open challenge for the best of the students, to give them an incentive for further improvement outside of the classroom. Table \ref{tab:complexity} lists the definition of stages.

\begin{table}
\centering
\caption{Different complexity stages}
\label{tab:complexity}
\begin{tabular}{llllll}
\hline\noalign{\smallskip}
Identification & Target\\
\noalign{\smallskip}
\hline
\noalign{\smallskip}
C1 & Easy to fix and identify & lecture\\
C2 & Target to reach for most of the students & practical work\\
C3 & Target to reach 100 percent in grading & practical work\\
C4 & Challenge for lecture & Extra work for deep knowledge\\
\hline
\end{tabular}
\end{table}

Stage one vulnerabilities({\em C1}) are easy to find, easy to exploit, and easy to fix. The vulnerabilities were shown in a very similar form in the lecture to show the concept of the vulnerability. Stage one vulnerabilities provide fast success and hence support a positive attitude towards BREW, motivate, and make eager to find vulnerabilities of the next stages.

Stage two vulnerabilities ({\em C2}) are designed for practical lessons and are suitable for the most common problems in IT-Security.
A stage two vulnerability is designed as a flaw without any side effect.
It can be found and fixed with best practice patterns presented in common textbooks.

Stage three vulnerabilities ({\em C3}) require a good understanding of IT security and web technologies to identify, explore, and fix. Stage three vulnerabilities combine different technologies and knowledge areas. Some built-in traps may decoy the solution. Stage three vulnerabilities are enhanced vulnerabilities that can be found out in the wild very often.

Stage four vulnerabilities ({\em C4}) are special challenges that are outside of the scope of bachelor or master lectures. They are designed as very sophisticated problems in IT-Security. A challenge is meant for individual learning or classroom contests. Some of these challenges are not solved by students yet. Usually, the best students of the course can find, exploit, and fix  at most one stage four vulnerability during a semester. Special care is taken that solutions of challenges do not leak to public. Until all challenges are mastered, a lecturer needs a personal identification to get the solution for the challenges from the BREW team.

\subsection{Educational Usage Scenarios}\label{subsec:usage}
BREW was not only designed to deal with heterogeneous hardware and software environments, it also was designed to be flexible in how it is used in teaching. Some of the possible usage scenarios are described in the following. Usage scenarios include White Box Testing, Black Box Testing, and  Hackveloper.

\subsubsection{White Box Testing Scenario} 
For White Box Testing, the student has access to the source code as well as to the running application. The scenario simulates a typical industrial quality testing in software development. Common used white box testing tools are already integrated into BREW. This includes tools for manual audits as well as tools integrated into the IDE like pattern matching or data flow analysis. BREW ensures success for different White Box Testing tools.

\subsubsection{Black Box Testing}
Beside White Box Testing, BREW can be used for Black Box Testing. Black Box Testing differs from White Box Testing in that the students do not have access to the source code of the application.  To do so, BREW can be deployed as pure server with the command line. It is designed to deal with common vulnerability testing tools (see table \ref{tab:ostools}). The Black Box Testing scenario simulates the work of a typical penetration tester.

\begin{table}
\centering
\caption{Examples for common black box testing tools}
\label{tab:ostools}
\begin{tabular}{llllll}
\hline\noalign{\smallskip}
Identification & Target\\
\noalign{\smallskip}
\hline
\noalign{\smallskip}
TS1 & BurpSuite\\
TS2 & Owasp Zap\\
TS3 & Metasploit\\
TS4 & nmap, dnsmap, hping3\\
\hline
\end{tabular}
\end{table}

\subsubsection{Hackveloper}
Neither the White Box Testing nor the Black Box Testing approach helps students to write secure code. The third scenario, called Hackveloper, is the envisioned use of BREW. Students are not only asked to identify and exploit vulnerabilities, they also must fix any vulnerability they find. In comparison to the other two approaches, students can find and fix vulnerabilities. Naming Implementation of functionality follow Best Practice Patterns similar to  \cite{meucci2008owasp,van2008owasp,vanderstock2014} and \cite{springdoc2014}. This ensures that students are able to find similar problems within common literature or online documentations.


\subsection{Architecture and General Implementation}
The architecture of the web application follows the MVC (Model-View-Controller) pattern.
The implementation of BREW uses the Spring \cite{spring2014} framework.
The Spring framework is a lightweight platform for java applications.

BREW uses Java as it is a well known, well documented and widely used programming language.
However most of the built-in vulnerabilities are independent from the underlying language.

\begin{figure}
\centering
\includegraphics[width=1\textwidth]{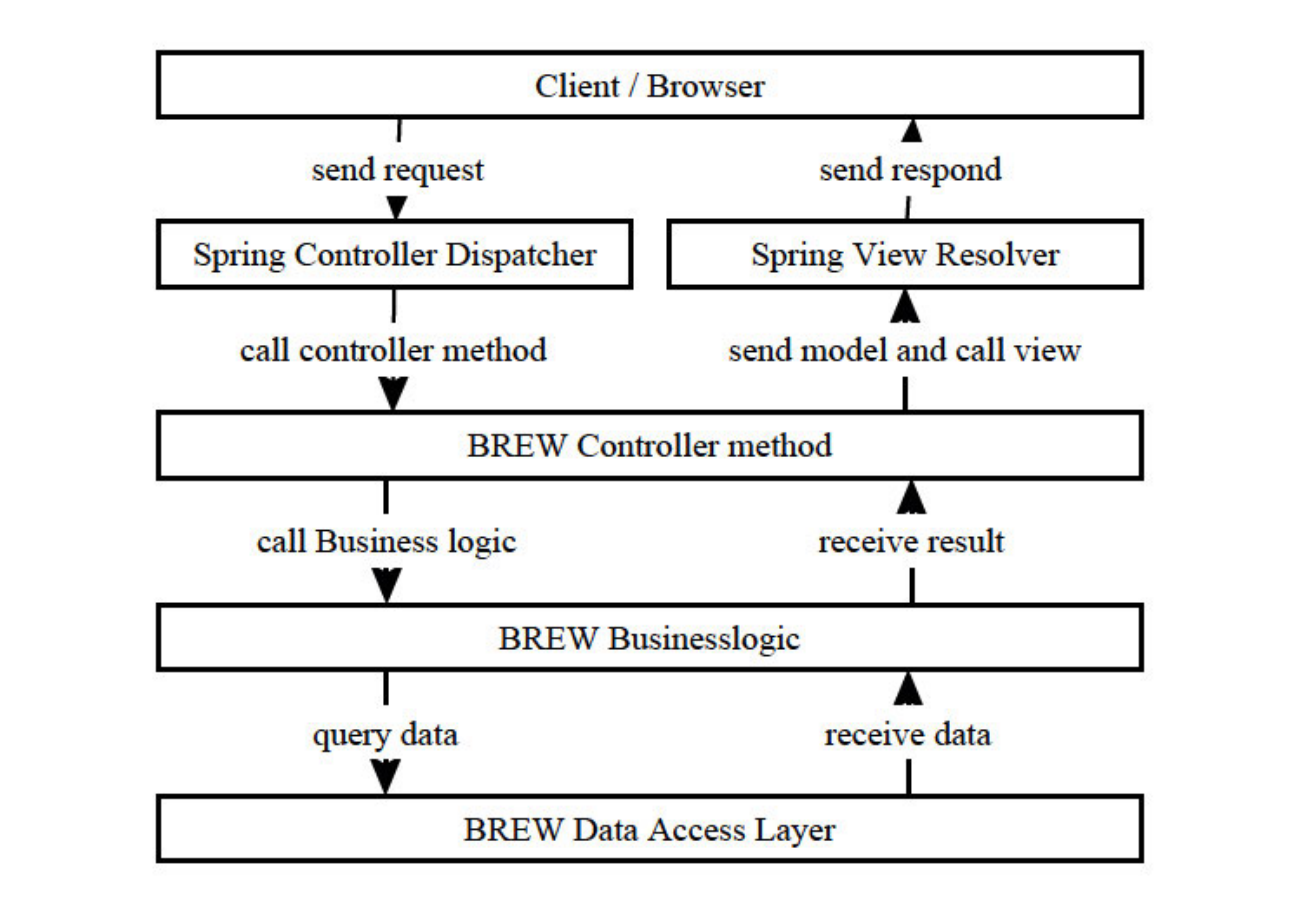}
\caption{BREW architecture} 
\label{fig:architecture}
\end{figure}

A schematic overview is given in Figure \ref{fig:architecture}.

At first, the browser sends a request to the web application. Within the path matching of the Spring framework a controller method gets called, see listing 1. In this example the search page is mapped to {\em http://mydomain/search.secu}. For simplicity every logical component has a related controller class (e.g., {\em Search page $\rightarrow$ SearchController.java}). Every action within this logical component inherits a method (p.Ex {\em search $\rightarrow$ searchWebWithPost($\dots$)}).
\\\\
\noindent
{\it Listing 1. Example for search controller }
\begin{small}
\begin{verbatim}
@RequestMapping(value = "/search.secu", method = RequestMethod.POST)
public ModelAndView searchWebWithPost(
   @RequestParam(value = "search", required = false) String search) {
   ModelAndView mv = new ModelAndView("search");
   mv.addObject("searchString", search);
   return mv;
}
\end{verbatim}
\end{small}

This method implements the business logic. It produces a HashMap-based model for further usage in the related view. See Listing 2 for the model. In this example, the view gets the variable {\em search} accessible with the key {\em searchString}.
\\\\
\noindent
{\it Listing 2. Injection of a model}
\begin{small}
\begin{verbatim}
mv.addObject("searchString", search);
\end{verbatim}
\end{small}

As shown in Listing 3 the view with the name {\em search} will be used.
In the case of BREW a view name automatically links to a java server page with the suffix {\em .jsp}. In this case it will be {\em search.jsp}.
\\\\
\noindent
{\it Listing 3. Creation of a view}
\begin{small}
\begin{verbatim}
ModelAndView mv = new ModelAndView("search");
...
return mv; 
\end{verbatim}
\end{small}

The given model ensures that students need to understand the dependency between the controller method and the related view to solve all stage one to stage three problems.
Understanding the Model-View-Controller pattern is very important for understanding the security of web applications.

\subsection{Vulnerabilities}
Vulnerability design in BREW follows common flaws in web applications.
These common flaws can be classified as given in table \ref{tab:vulntype}.

\begin{table}
\centering
\caption{Different vulnerability types}
\label{tab:vulntype}
\begin{tabular}{llllll}
\hline\noalign{\smallskip}
Identification & Vulnerability type\\
\noalign{\smallskip}
\hline
\noalign{\smallskip}
F1 & Bad programming style\\
F2 & Design and architecture flaws\\
F3 & Wrong implementation of basic concepts\\
F4 & Vulnerable environment\\
\hline
\end{tabular}
\end{table}

For example, a bad programming style {\em F1} can be found in an  SQL injection vulnerability (stage two vulnerability). Listing 4 shows an example of the vulnerability. To make the problem harder, most of the SQL query is validated as proposed as best practice in the associated lecture. However, string concatenation is used for the last variable {\em uid} as a programmer that does not understand the problem of SQL injection would probably do. Concatenating uid to the SQL query allows a successful SQL injection. As some of the variables of the SQL query are correctly validated, finding the vulnerability is made more difficult.

An attacker can dispose any valid sql-code over this variable.
\\\\
\noindent
{\it Listing 4. Example for bad programming style }
\begin{small}
\begin{verbatim}
String sql = "update M_USER set " +
             "muname = ?, " +
             "mpwd = ? " +
             "where " +
             "ID = "+uid;

jdbcTemplate.update(sql, 
	new Object[]{uname, upwd}, 
	new int[]{Types.VARCHAR, Types.VARCHAR});
}
\end{verbatim}
\end{small}

Design and architecture flaws {\em F2} are introduced into source code by bad planning. One example for a design flaw in BREW are data access functionalities: there is no single point of validation but validation has to be implemented on multiple points for each function in BREW. This is very bad design. Vulnerabilities are introduced into BREW by forgetting some validations on purpose.

The third flaw is wrong implementation of basic concepts {\em F3}. BREW implements a buggy storage of passwords. The common concept of password storage is to use salted hashes. In BREW, the admin password is secured without a salt and with the weak md5 algorithm. The password hash can be found in a rainbow table as the admin uses a bad password. In this case a simple google query results to the origin password.

Another class of flaws is the use of weaknesses of used frameworks {\em F4}.
An example for a vulnerability of this class in BREW is that the management platform of the embedded tomcat uses the default password. Within this vulnerability an attacker can take over the full application.
\subsection{Extensibility}
BREW is an open source project.
The first release will be able at github.
At the moment it is a release candidate with a few minor fixation needed.
It is planned to build an open community to extend BREW.

The current version of BREW implements common functionalities for a web application.
Hence BREW is designed to work with only less configuration overhead it is easy to extend.
A new functionality, vulnerability or page can be integrated by writing a new class or bunch of classes.
The registration of an extension is done with spring annotations.

However it is also possible to build new web services by extending the configuration of the built-in Tomcat server.

BREW gets compiled whenever restarting the webserver.
New libraries, source files or binaries are inherited when present under the main BREW directory tree.

To extend a lecture, a developer just needs to copy the new files to BREW.
A makefile in the root directory can be used to pack the new BREW distribution.
This functionality is explained in detail in the developer handbook.
\section{Deployment}\label{sec:deploy}
BREW was designed to deal with heterogeneous hardware and software environments. Some of the possible deployment scenarios are described in the following. 

\subsection{Vulnerability Selection}
BREW has a modular design.
Lecturers can deploy BREW with selected vulnerabilities or with the complete set.
In all cases BREW can be installed by unpacking a single file.
To fulfil different requirements for different lectures, the lecturer can build an adapted version of the application for each lecture or even different groups of students.
Selection of vulnerabilities and functionalities for the final BREW deployment is possible by using compile flags of the makefile.
However, complex dependencies may exist between stage three and stage four vulnerabilities.
Table \ref{tab:conflevel} describes dependencies and the resulting constraints in the selection of vulnerabilities.

\begin{table}
\centering
\caption{Configuration level and their dependencies}
\label{tab:conflevel}
\begin{tabular}{llllll}
\hline\noalign{\smallskip}
Identification & Dependency\\
\noalign{\smallskip}
\hline
\noalign{\smallskip}
$V_{C1}$ & $vc1_1 \vee \cdots \vee vc1_n$\\
$V_{C2}$ & $V_{C1} \wedge (vc2_1 \vee vc2_n)$\\
C3 & ($V_{C1} \wedge V_{C2}$)\\
C4 & Challenge for lecture & Extra work for deep knowledge\\
\hline
\end{tabular}
\end{table}
Within the stages {\em C1} and {\em C2} every Vulnerability can be selected or unselected. The stage {\em C3} needs all vulnerabilities from {\em C1 - C2}. Whenever {\em C4} is selected, BREW gets build in the full version.

\subsection{The Lecturer's Cheat Sheet}
Each vulnerability of BREW is well documented. A reference solution on how to fix each vulnerability is also provided for the lecturers. BREW also comes with python scripts for easy demonstration of exploits. These scripts provide an easy way to show the reference solution during a lecture. The scripts may also be used to validate the fixed code of the students. Code fixes are proposed as patches. With these patches the solution can be explained and ensures that every student has the same version of the source code for following lectures.

\subsection{BREW as IDE project}
Originally, BREW is deployed as IDE project. A configuration is prepared for the eclipse IDE and IntelliJ. The IDE project includes every dependency and a start configuration. This approach allows BREW to be started by clicking the run button in the IDE. Whenever the source code of BREW gets changed, the user needs to restart the project to deploy his changes. Even a hot deployment while debugging is possible in most cases. (Expect for {\em C4} challenges)
The project path includes a fully qualified git-tree. To update BREW or to get a clean version or a version with patches, the user can pull different versions from the remote git repository. 

\subsection{Local Virtual Machine Deployment and Cloud Deployment}
To ensure that every student has the same environment, BREW is available as virtual machine. For use on student laptops or on lab computers, a Virtual Box image is used. For deployment in a cloud environment like Cloudstack or OpenStack, a version for qemu-KVM exists. Future versions of BREW will also be compliant to Amazon Cloud and VMWare ESXI \cite{vmware2014}. Using BREW in the cloud ensures a fast deployment for large student groups without the overhead of local installation. 

\subsection{BREW as Target Server}
Without an IDE, BREW can be used as a pure server for Black Box Testing as well as a vulnerable target in IT-security research. The server consists of BREW and an optional reverse proxy for performance evaluation.
With a plain installation of Kali Linux in another virtual machine, students have a full set of penetration tools and the target.
This version is also available for Virtual Box and qemu-KVM.

\section{Experiences with BREW}\label{sec:exp}
Since today, BREW has been tested in different lectures and practical lessons.
\begin{itemize}
\item lectures in IT-Security (MUAS)
\item lectures in IT-Forensics (MUAS)
\item lectures in Intrusion Detection (MUAS)
\item Developer Days 2013 (Munich)
\item Erasmus IP Program 2012 (MUAS)
\item Erasmus IP Program 2013 (University of South Wales)
\item Erasmus IP Program 2014 (University of Santander)
\item DOAG Conference 2013 (Nuremberg)
\end{itemize}

\section{Conclusion}\label{sec:conclusion}
This paper presented BREW, a flexible, platform-independent tool for teaching secure programming of web applications. BREW takes an Hackveloper approach in which students not only have to identify and exploit vulnerabilities of web applications, but they also have to fix and test the code of a vulnerable web application. However, BREW can also be used in many other teaching scenarios, e.g., in teaching Black Box Testing or White Box Testing. BREW has been tested in national and international teaching settings and proved to be quite useful. 
In the future, BREW will be continuously be extended with new vulnerabilities to stay up to date with state of the art of attacks on web applications. BREW vulnerabilities will implement vulnerabilities for further versions of the OWASP Top Ten project as well as vulnerabilities of other similar lists. 

\bibliographystyle{splncs}

\bibliography{bib.bib}

\end{document}